# Understanding of collective atom phase control in modified photon echoes for a near perfect, storage time extended quantum memory

Rahmatullah[1,2] and B. S. Ham[1†]
[1]Center for Photon Information Processing, and School of Electrical Engineering and Computer Science, Gwangju Institute of Science and Technology, Gwangju, 61005 Republic of Korea
[2]Quantum Optics Lab., Department of Physics, COMSATS Institute of Information Technology, Islamabad, Pakistan

†Corresponding author: bham@gist.ac.kr

**Abstract:** A near perfect, storage time-extended photon echo-based quantum memory protocol has been analyzed by solving the Maxwell–Bloch equations for a backward scheme in a three-level system. The backward photon echo scheme is combined with a controlled coherence conversion process via control Rabi flopping to a third state, where the control Rabi flopping collectively shifts the phase of the ensemble coherence. The propagation direction of photon echoes is coherently determined by the phase matching condition between the data (quantum) and the control (classical) pulses. Herein we discuss the classical controllability of a quantum state for both phase and propagation direction by manipulating the control pulses in both single and double rephasing photon echo schemes of a three-level system. Compared with their well-understood use for two-level photon echoes, the Maxwell–Bloch equations to a three-level system have a critical limitation regarding the phase change when interacted with an arbitrary control pulse area.

**Introduction**

Over the last decade modified photon echo techniques have been intensively studied and applied to quantum memory applications to overcome the fundamental limitation of population inversion in conventional photon echoes, for which the population inversion excited by an optical $\pi$ pulse results in quantum noises caused by spontaneous and/or stimulated emissions [1-17]. Although some of these techniques have been successful for quantum state storage and retrieval [8-13], the understanding of collective atom phase control has still been limited, where absorptive photon echoes have been involved in controlled atomic frequency comb (AFC) echoes [8,9] and doubly rephased (DR) photon echoes [10-13]. To coherently manipulate the absorptive photon echoes, a controlled coherence conversion (CCC) process has been proposed [18] and experimentally demonstrated [19]. Recent observations of photon echoes in single [8,9] and double [10-13] rephasing photon echo schemes, however, do not contradict the CCC theory, but reveal coherence leakage by Gaussian light pulses, resulting in echo generations all the time regardless of the pulse area whose maximum efficiency reaches as high as 26% [20]. In addition to our previous studies on numerical [14-16] and analytical [17] approaches, herein we discuss the CCC theory by using the commonly applied Maxwell–Bloch (MB) approach to correct the critical mistakes in previous works [8-13] and thus to contribute to the implementation of photon echo–based quantum memory.

In the study of modified photon echo-based quantum memories, using MB equations has been a common theoretical tool. The MB equations have the advantage of using both space and time variables, and thus are practical for calculations of photon echo retrieval efficiency with respect to the optical depth (or physical length) of an ensemble medium [1-4]. However, the MB approach is completely unable to give exact solutions of individual atom phase evolutions, and thus continuous tracing of ensemble coherence change in the time domain is completely impossible. This limitation requires us to totally rely upon inevitable assumptions for the atom-field interactions such as rephasing and CCC. In other words the MB theory prevents us from obtaining exact answers to the phase change of the ensemble coherence with respect to the pulse area of an interacting optical field.

To overcome this limitation of the MB approach, we have dealt with full numerical [14-16] and full analytical [17] solutions based on time-dependent density matrix equations to exactly trace the coherence evolutions of an ensemble. As a result, we have introduced the controlled double rephasing (CDR) echo protocol [14] based upon CCC [18]. Although CDR is perfect to investigate temporal coherence behaviors of coherent transient phenomena, such techniques of numerical and analytical calculations are limited in optical depth-dependent photon echo efficiency. In the present Report, we comply with the MB equations, firstly to confirm the CDR echo protocol with previous results of CCC in Refs. 17 and 18, secondly to analyze the critical mistakes in Refs. 8-13 due to incorrect assumptions, and finally to discuss ensemble phase evolutions and their phase control in both single [8,9] and double rephasing schemes [10-14]. Near perfect retrieval efficiency in quantum memories is critical for both fault-tolerant quantum computing [21] and loophole-free quantum cryptography [22]. Thus the present Report of the near perfect, storage-time extended quantum memory protocol should shed lights on the future quantum information area using quantum memories.

## Theory: Maxwell–Bloch equations
### A. Conventional two-pulse photon echo

We consider a three-level optical ensemble medium composed of $N$ indistinguishable atoms. The energy level diagram and the pulse sequence of the present CDR echo scheme are depicted in Fig. 1. The data (D), the first rephasing ($R_1$) and the second rephasing ($R_2$) pulses are resonant to the transition of $|1\rangle - |2\rangle$ to satisfy the requirements of the DR photon echo scheme [10-17]. The control pulses $C_1$ and $C_2$ are resonant between states $|2\rangle$ and $|3\rangle$ to enable CCC [14-19]. Initially all atoms of the medium are in the ground state, $|1\rangle$. For an ideal system, all decay rates are neglected, unless specified otherwise. All light pulses are collinear (or near collinear) in the z-axis. To make the first echo ($E_1$) silent to avoid affecting the final echo ($E_2$), both rephasing pulse propagation directions are set to be opposite (backward) to that of the D pulse [10]. To satisfy the backward photon echo condition, the control pulses are set to be counter-propagating each other [19]: $\vec{k}_{E_2} = \vec{k}_{C_1} + \vec{k}_{C_2} - \vec{k}_D$; each pulse $j$ is characterized by a wave vector $\vec{k}_j$. Unlike phase mismatching in silent echo ($E_1$) formation [10], which is determined by D and $R_1$, the final echo ($E_2$) propagation direction ($\vec{k}_{E_2}$) is determined by the control pulses [14]. The MB equations for the atomic coherence and the D pulse are respectively denoted as follows:

$$\frac{\partial}{\partial t}\sigma_{12}(z,t,\Delta) = i\Delta\sigma_{12}(z,t,\Delta) + i\varepsilon_D(z,t), \tag{1}$$

$$\frac{\partial}{\partial z}\varepsilon_D(z,t,\Delta) = \frac{i\alpha}{2\pi}\int_{-\infty}^{\infty}\sigma_{12}(z,t,\Delta)d\Delta, \tag{2}$$

where $\alpha$ is the optical depth parameter, and $\Delta$ is the detuning of the atom (see Supplementary Information). The ensemble is inhomogeneously broadened by $\Delta' = \sum_j \Delta_j$. Here, we consider the case of a $\Delta$–detuned atom for simplicity. Thus, the solution for atomic coherence is given by:

$$\sigma_{12}(z,t,\Delta) = i\int_{-\infty}^{t}\varepsilon_D(z,\acute{t})e^{i\Delta(t-\acute{t})}d\acute{t}. \tag{3}$$

The positive sign of this equation represents that the atomic coherence excited by the D pulse is absorptive. It should be noted that $\sigma_{12}(z,t,\Delta) = -\rho_{12}(z,t,\Delta)$, where $\rho_{12}(z,t,\Delta)$ is the density matrix element. The solution of equation (2) gives the well-known Beer's law: $\varepsilon_D(z,t) = \varepsilon_D(0,t)e^{-\alpha z/2}$. The D pulse is assumed to be fully absorbed by the medium, thereby transferring its quantum information (phase, amplitude, polarization, etc.) into the collective atomic coherence. We assume that the D pulse is very weak to treat it as a quantum state. Thus, the ground state population change arising from the D pulse excitation is neglected: $\sigma_{11}(z,t,\Delta) = 1$ and $\sigma_{22}(z,t,\Delta) = 0$.

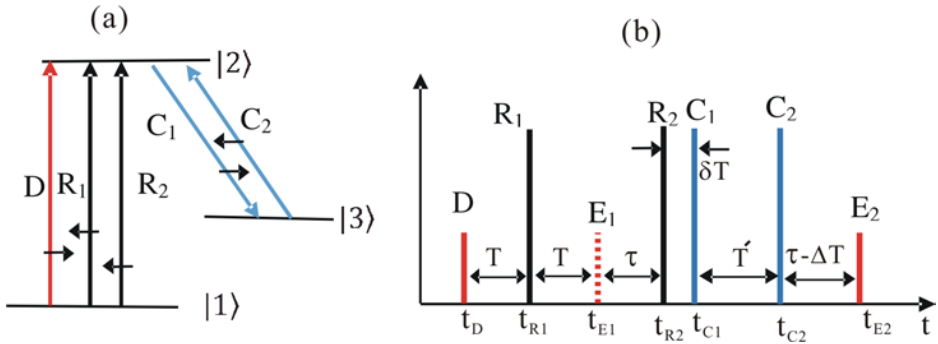

**Figure 1. Schematics of controlled double rephasing (CDR) echoes.** (a) Energy level diagram for CDR echo. The short black arrows indicate the pulse propagation direction. (b) Pulse sequence for (a), where $t_j$ is the arrival time of pulse j.

With the time delay of T as shown in Fig. 1b, the rephasing pulse ($R_1$) arrives at $t = t_{R_1}$. The $R_1$ pulse has a π pulse area, and its propagation direction is opposite to that of the D pulse ($\vec{k}_{R_1} = -\vec{k}_D$). Here the function of the $R_1$ pulse is to swap the populations between states $|1\rangle$ and $|2\rangle$: $\sigma_{11}(z, t_D, \Delta) \overset{R_1}{\leftrightarrow} \sigma_{22}(z, t_{R_1}, \Delta)$. With the weak D pulse, the swapping result is $\sigma_{22}(z, t, \Delta) = 1$ and $\sigma_{11}(z, t, \Delta) = 0$. Thus, the atomic coherence arising from the application of the $R_1$ pulse becomes:

$$\frac{\partial}{\partial t}\sigma_{12}(z, t, \Delta) = i\Delta\sigma_{12}(z, t) - i\varepsilon_{R_1}(z, t). \tag{4}$$

The solution of equation (4) is

$$\sigma_{12}(z, t, \Delta) = e^{i\Delta(t - t_{R_1})}\sigma_{12}(z, t_{R_1}, \Delta) - i\int_{t_{R_1}}^{t}\varepsilon_{R_1}(z, \acute{t})e^{i\Delta(t - \acute{t})}d\acute{t}. \tag{5}$$

Because the π pulse reverts the coherence evolution direction, $\sigma_{12}(z, t_{R_1}, \Delta)$ is equal to the conjugate of equation (3) at $t = t_{R_1}$, namely $\sigma_{12}(z, t_{R_1}, \Delta) = [\sigma_{12}(z, t = t_{R_1}, \Delta)]^\dagger$. Therefore, equation (5) can be rewritten as

$$\sigma_{12}(z, t, \Delta) = -i e^{-i\Delta(2t_{R_1} - t)}\int_{-\infty}^{\infty}\varepsilon_D^\dagger(z, \acute{t})e^{i\Delta\acute{t}}d\acute{t} - i\int_{t_{R_1}}^{t}\varepsilon_{R_1}(z, \acute{t})e^{i\Delta(t - \acute{t})}d\acute{t}, \tag{6}$$

where the first term represents free evolution and the second term represents interaction with the rephasing pulse. The ensemble coherence generated by the D pulse becomes in phase and results in a photon echo $E_1$ at $t = t_{E_1} = 2t_{R_1} - t_D$. The propagation direction of the first echo $E_1$ determined by the first rephasing pulse $R_1$ is $\vec{k}_{E_1} = 2\vec{k}_{R_1} - \vec{k}_D = -3\vec{k}_D$. Due to the phase mismatch between D and $E_1$, the echo signal (macroscopic coherence) cannot be generated due to complete out of phase [10]: Silent echo.

### B.  DR echo

To rephase the system one more time, the second rephasing ($R_2$) π pulse is followed by $E_1$ to satisfy the requirements of the DR photon echo scheme. In the DR scheme, the excited state population after the final echo $E_2$ is the same as that after the D pulse excitation. This means that all excited-state atoms in the DR scheme should contribute to the echo signal without contributing to quantum noises. The MB equation for DR is similar to that for D, and the optical coherence solution for the final echo $E_2$ is

$$\sigma_{12}(z, t, \Delta) = e^{i\Delta(t - t_{R_2})}\sigma_{12}(z, t_{R_2}, \Delta) + i\int_{t_{R_2}}^{t}\varepsilon_{R_2}(z, \acute{t})e^{i\Delta(t - \acute{t})}d\acute{t}, \tag{7}$$

where $\sigma_{12}(z, t_{R_2}, \Delta)$ is equal to the conjugate of equation (6) at $t = t_{R_2}$:

$$\sigma_{12}(z, t, \Delta) =$$
$$i e^{-i\Delta(2t_{R_2} - 2t_{R_1} - t)}\int_{-\infty}^{\infty}\varepsilon_D(z, \acute{t})e^{-i\Delta\acute{t}}d\acute{t} + i e^{-2i\Delta t_{R_2}}\int_{-\infty}^{\infty}\varepsilon_{t_{R_1}}^\dagger(z, \acute{t})e^{i\Delta(t + \acute{t})}d\acute{t} + i\int_{t_{R_2}}^{t}\varepsilon_{R_2}(z, \acute{t})e^{i\Delta(t - \acute{t})}d\acute{t}. \tag{8}$$

The final echo $E_2$ is formed at $t = t_{E_2}$ as the rephasing result of $E_1$ by $R_2$, and its propagation direction is forward if $\vec{k}_{R_2} = \vec{k}_{R_1}$: $\vec{k}_{E_2} = 2\vec{k}_{R_2} - \vec{k}_{E_1} = 2\vec{k}_{R_2} - (2\vec{k}_{R_1} - \vec{k}_D) = \vec{k}_D$. However, the retrieved signal $E_2$ is absorbed by the medium due to absorptive coherence as shown in equation (8). In other words, echo $E_2$ cannot be radiated from the medium. This fact has already been explored numerically [14-16] and analytically [17]. From now on we discuss and correct critical mistakes in previous analyses [8-13]. By the way, the observation of DR echoes [10-13] has been understood as a coherence leakage due to an imperfect rephasing process by Gaussian-distributed light in a transverse mode, resulting in the leakage-caused maximum retrieval efficiency at 26% [20].

**Discussion**

### A. CDR echo

To fix the absorptive echo problem in DR, the CDR echo protocol has been proposed whereby the control pulse pair $C_1$ and $C_2$ are added as shown in Fig. 1b [14]. As discussed in Refs. [14-18], the function of the control pulses is not only to convert the coherence between optical and spin states via population transfer, but also to induce a collective phase shift. Unlike the DR scheme of equation (8), the propagation vector of $E_2$ can be controlled to be backward if $\vec{k}_{C_1} = -\vec{k}_{C_2}$: $\vec{k}_{E_2} = \vec{k}_{C_1} + \vec{k}_{C_2} - \vec{k}_D = -\vec{k}_D$ [14,19]. Here, the rephasing pulses have nothing to do with the phase matching for $\vec{k}_{E_2}$. Unlike the forward echo in the conventional two-pulse photon echo scheme, which suffers from reabsorption by noninteracting atoms according to Beer's law, the backward echo $E_2$ is free from reabsorption, resulting in near perfect echo efficiency [1,2,19]. To eliminate any potential two-photon coherence between states |1⟩ and |3⟩, the $C_1$ pulse is delayed by $\delta T$ from $R_1$. Here, it should be noted that the *macroscopic* two-photon coherence is sustained only within the overall optical coherence time determined by the inverse of the atom broadening $\Delta'$. However, *individual* atom coherence is sustained regardless of atom broadening for the optical phase decay time $T_2$, where $T_2 \gg \Delta T$ [18]. Here the D pulse duration is practically comparable to (or a bit longer than) $1/\Delta'$. Thus, by simply neglecting $\delta T$, the atomic coherence at $t_{C_1}$ can be expressed as:

$$\sigma_{12}(z, t_{C_1}, \Delta) = i\, e^{-i\Delta(2t_{R_2} - 2t_{R_1} - t_{C_1})} \int_{-\infty}^{\infty} \varepsilon_D(z, \acute{t}) e^{-i\Delta \acute{t}} d\acute{t}. \tag{9}$$

In equation (9) we have only considered the first term of equation (8), which is related to the free evolution of the coherences generated by the D pulse; the second and third terms are associated with the rephasing fields. Because the first echo $E_1$ is silent, and the second echo $E_2$ is not emitted owing to its absorptive coherence, we can remove the last two terms of equation (8). With a $\pi$ $C_1$ pulse, optical and spin coherence respectively satisfy the following relations: $\sigma_{12}(z, t, \Delta) = \cos\left(\frac{\pi}{2}\right) \sigma_{12}(z, t_{C_1}, \Delta) = 0$; $\sigma_{13}(z, t, \Delta) = e^{i\pi/2} \sigma_{12}(z, t_{C_1}, \Delta)$ (see Refs. 14-18). The $C_1$ pulse induces a $\pi/2$ phase shift and locks the optical coherence until $C_2$ arrives. Here, the novel property of the $\pi/2$ phase shift by the $\pi$ pulse area of $C_1$ (or $C_2$) originates in the Raman (two-photon) coherence, where a $2\pi$ Raman pulse in a three-level system actually plays as a $\pi$ pulse does in a two-level system: see Figs. 3 and 4 of Ref. [23] and Fig. 4 of Ref. [24]. There is no way to expect this novel property from the MB approach. Without *a priori* understanding of the coherence behavior in a three-level system, the same mistake has been repeated in the controlled AFC echoes [8,9]; this will be discussed further in section B.

The atomic coherence after the $C_2$ pulse is

$$\frac{\partial}{\partial t}\sigma_{12}(z, t, \Delta) = i\Delta \sigma_{12}(z, t, \Delta). \tag{10}$$

Equation (10) is obtained by substituting $\varepsilon_l = 0$ and $\sigma_{13}(z, t, \Delta) = 0$ (no spin coherence after $C_2$) into equation (S4) of the Supplementary Information. The solution of equation (10) is

$$\sigma_{12}(z, t, \Delta) = \sigma_{12}(z, t_{C_2}, \Delta) e^{i\Delta(t - t_{C_2})}. \tag{11}$$

The $C_2$ pulse also brings a $\pi/2$ phase shift while swapping the spin and optical coherence:

$$\sigma_{12}(z, t_{C_2}, \Delta) = e^{i\pi/2} \sigma_{13}(z, t, \Delta) = e^{i\pi} \sigma_{12}(z, t_{C_1}, \Delta) = -i\, e^{-i\Delta(2t_{R_2} - 2t_{R_1} - t_{C_1})} \int_{-\infty}^{\infty} \varepsilon_D(z, \acute{t}) e^{-i\Delta \acute{t}} d\acute{t}. \tag{12}$$

Substituting equation (12) into equation (11) gives

$$\sigma_{12}(z, t, \Delta) = -i\, e^{-i\Delta(t_{C_2} - t_{C_1} + 2t_{R_2} - 2t_{R_1} - t)} \int_{-\infty}^{\infty} \varepsilon_D(z, \acute{t}) e^{-i\Delta \acute{t}} d\acute{t}, \tag{13}$$

where the negative sign represents that the echo $E_2$ is emissive due to the $\pi$ phase shift induced by the control pulse pair, which has already been extensively explored numerically [14-16] and analytically [17]. Thus, the echo $E_2$ propagates backward without reabsorption and is radiated out of the medium with near perfect retrieval efficiency; this will be discussed further in Fig. 2.

### B. Single rephased photon echo

Now we consider the coherence evolution for a controlled single rephasing scheme with $R_2=0$ in Fig. 1 [18,19]. This scheme itself cannot be used for quantum memory because the population inversion has not been solved yet. However, this scheme itself implies the AFC echo scheme, whereby all excited-state atoms freely decay into a dump state. The AFC scheme can be easily obtained by swapping the pulse sequence of D and $R_1$, and substituting $R_1$ with a repeated weak two-pulse train (see Supplementary Information). So, neglecting the population constraint issue without $R_2$, the final optical coherence can be obtained by applying the control pulses to equation (6) as follows (see Section III of the Supplementary information):

$$\sigma_{12}(z,t,\Delta) = i\, e^{-i\Delta(t_{C_2}-t_{C_1}+2t_{R_1}-t)} \int_{-\infty}^{\infty} \varepsilon_D^\dagger(z,\acute{t})e^{i\Delta \acute{t}}\, d\acute{t}. \tag{14}$$

The positive sign of this equation represents the absorptive coherence, which is the same problem as in the DR scheme in equation (8). In other words, the π–π control pulse sequence added to the single rephasing scheme inverts the system coherence to be absorptive [18]. To convert the absorptive echo into an emissive one, simply one more control Rabi flopping is added, for which the $C_2$ pulse area is increased to $3\pi$, i.e., $\sigma_{12}(z,t,\Delta) \xrightarrow{C_1(\pi)} e^{-\left(\frac{\pi}{2}\right)i}\sigma_{13}(z,t,\Delta) \xrightarrow{C_2(\pi)} e^{\pi i}\sigma_{12}(z,t,\Delta) \xrightarrow{C_2(\pi)} e^{-\left(\frac{3\pi}{2}\right)i}\sigma_{13}(z,t,\Delta) \xrightarrow{C_2(\pi)} e^{2\pi i}\sigma_{12}(z,t,\Delta)$, and thus $\sigma_{12}(z,t,\Delta)$ is negative. This means that the π–π control pulse sequence in Refs. [8] and [9] must be corrected to be a π–3π control pulse sequence. Here, the π–π control pulse sequence in Ref. [1] is using a Doppler effect–caused π-phase shift, resulting in an emissive echo. This is not the case for a solid medium [2,14,18]. The observation of controlled AFC echoes with a π–π control pulse set [8,9], however, is due to the imperfect rephasing by commercial light sources with Gaussian spatial distributions [20]. The coherence leakage in a DR scheme induced by the Gaussian pulse limits the echo efficiency to ~10% on average regardless of the rephasing pulse area, whereas its maximum efficiency reaches 26% for a π/2 pulse area in a DR scheme [20].

### C. Near perfect retrieval efficiency in CDR

We now calculate the echo efficiency of the present CDR echo scheme shown in Fig. 1. Because the echo $E_2$ is emitted in the backward direction, the atomic coherence and optical field in the backward direction can be respectively represented as follows [25]:

$$\frac{\partial}{\partial t}[\sigma_{12}(z,t,\Delta)]_b = i\Delta[\sigma_{12}(z,t,\Delta)]_b + i\varepsilon_b(z,t), \tag{15}$$

$$\frac{\partial}{\partial z}\varepsilon_b(z,t) = -\frac{i\alpha}{2\pi}\int_{-\infty}^{\infty}[\sigma_{12}(z,t,\Delta)]_b d\Delta. \tag{16}$$

The solution of equation (15) is

$$[\sigma_{12}(z,t,\Delta)]_b = [\sigma_{12}(z,0,\Delta)]_b e^{i\Delta t} + i\int_0^t \varepsilon_b(z,\acute{t})e^{i\Delta(t-\acute{t})}\, d\acute{t}, \tag{17}$$

where $[\sigma_{12}(z,0,\Delta)]_b$ is obtained by setting $t=0$ in equation (13).

$$[\sigma_{12}(z,0,\Delta)]_b = \sigma_{12}(z,t=0,\Delta) = -i\, e^{-i\Delta(t_{C_2}-t_{C_1}+2t_{R_2}-2t_{R_1})} \int_{-\infty}^{\infty} \varepsilon_D(z,\acute{t})e^{-i\Delta \acute{t}}\, d\acute{t}. \tag{18}$$

Inserting equation (18) into equation (17) gives:

$$[\sigma_{12}(z,t,\Delta)]_b = -i\, e^{-i\Delta(t_{C_2}-t_{C_1}+2t_{R_2}-2t_{R_1}-t)} \int_{-\infty}^{\infty} \varepsilon_D(z,\acute{t})e^{-i\Delta \acute{t}}\, d\acute{t} + i\int_0^t \varepsilon_b(z,\acute{t})e^{i\Delta(t-\acute{t})}\, d\acute{t}. \tag{19}$$

Equations (19) and (16) are solved by means of Laplace transformation. For the backward scheme, $\varepsilon_b(L,\omega)=0$ because there is no field at $z=L$. Assuming an ideal case of complete absorption at $z=L$, solving equations (19) and (16) yields:

$$\varepsilon_b(0,\omega) = (1-e^{-\alpha L})e^{-i\omega(t_{C_2}-t_{C_1}+2t_{R_2}-2t_{R_1}-t)}\varepsilon_D(0,\omega). \tag{20}$$

The inverse Fourier transform gives:

$$\varepsilon_b(0,t) = (1 - e^{-\alpha L})\varepsilon_D(0, t - (t_{C_2} - t_{C_1} + 2(t_{R_2} - t_{R_1}))), \tag{21}$$

where the echo is emitted at $t = t_{E_2} = t_{C_2} - t_{C_1} + 2(t_{R_2} - t_{R_1}) + t_D$. Because $t_{C_2} - t_{C_1}$ can be lengthened to be much longer than $t_{R_1} - t_D$ in a Zeeman scheme [26], an additional but important function of the control pulse set is storage time extension [8,9,19]. The efficiency of the final backward echo $E_2$ is given by:

$$\eta = (1 - e^{-\alpha L})^2, \tag{22}$$

where $\eta$ can be near unity in an optically dense medium as shown in Fig. 2.

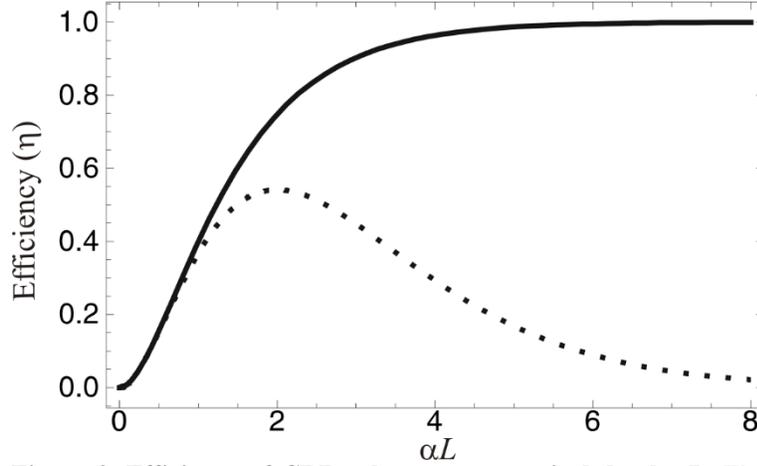

**Figure 2. Efficiency of CDR echoes versus optical depth $\alpha L$.** The solid curve is for the backward echo $E_2$; see Eq. (22). The dotted curve is for the forward echo $E_2$: $(\alpha L)^2 e^{-\alpha L}$.

On the other hand, the forward echo $E_2$ can be obtained by setting $\vec{k}_{C_1} = \vec{k}_{C_2}$, where $\vec{k}_{C_1} = \vec{k}_D$: $\vec{k}_{E_2} = \vec{k}_{C_1} + \vec{k}_{C_2} - \vec{k}_D = 2\vec{k}_{C_1} - \vec{k}_D = \vec{k}_D$. Here, the difference frequency between $C_1$ and D is just ~10 MHz, which is about $10^{-8}$ times the frequency of D in a rare-earth doped solid. The atomic coherence and the forward field satisfy the following relations:

$$\frac{\partial}{\partial t}[\sigma_{12}(z,t,\Delta)]_f = i\Delta[\sigma_{12}(z,t,\Delta)]_f + i\varepsilon_f(z,t), \tag{23}$$

$$\frac{\partial}{\partial z}\varepsilon_f(z,t) = \frac{i\alpha}{2\pi}\int_{-\infty}^{\infty}[\sigma_{12}(z,t,\Delta)]_f d\Delta. \tag{24}$$

By following the same procedure as used for the backward echo above, the optical field propagating in the forward direction can be expressed as follows:

$$\varepsilon_f(L,t) = \alpha L e^{-\frac{\alpha L}{2}}\varepsilon_D(0, t - (t_{C_2} - t_{C_1} + 2(t_{R_2} - t_{R_1}))). \tag{25}$$

Thus, the retrieval efficiency of the forward echo $E_2$ is $(\alpha L)^2 e^{-\alpha L}$ (see the dotted curve in Fig. 2). Obviously the forward echo E2 suffers from reabsorption.

In Fig. 2, we plot the echo efficiency of both the forward (dotted) and backward (solid) echo schemes for Fig. 1. For the forward echo, the retrieval efficiency reaches 54% at maximum, and then exponentially decreases due to the reabsorption process. On the other hand, the echo efficiency for the backward scheme approaches unity for $\alpha L \gg 0$ as expected according to Ref. 1 and as demonstrated in Ref. 19.

**Conclusions**

We analyzed and discussed the modified photon echo schemes of controlled double rephasing (CDR) for near perfect echo efficiency with no population inversion. We also pointed out the absorptive echo problems in previously demonstrated single (AFC) and double rephasing photon echo schemes, where the mistake in the controlled AFC echoes arises from a wrong assumption applied to the Maxwell–Bloch approach for a three-level system. Compared with a two-level system whose coherence harmonic is $2\pi$-based, the three-level system shows $4\pi$-based coherence harmonics. The Maxwell–Bloch method never gives an exact solution to the phase change of the ensemble coherence in its interaction with an arbitrary optical pulse area, unless known *a priori* as in the two-level system. With the help of full numerical [14-16] and analytic [17] calculations using time-dependent density matrix equations, our present Maxwell–Bloch calculation results confirmed the CDR echo theory. Thus, the present Report gives a clear understanding of collective atom phase control in modified photon echoes for quantum memory applications. The controlled coherence conversion process via control Rabi flopping to a third state was also investigated as a means of coherence inversion in a solid medium, whereby the absorptive coherence of the photon echo in the double rephasing scheme was converted into an emissive one. Finally, we discussed the near perfect CDR echo efficiency in a backward scheme using counter-propagating control Rabi pulses.

___________________________________________________________________________

**References**
1. Moiseev, S. A. and Kröll, S. complete reconstruction of the quantum state of a single-photon wave packet absorbed by a Doppler-broadened transition. *Phys. Rev. Lett.* **87**, 173601 (2001).
2. Moiseev, S. A. Tarasov, V. F. and Ham, B. S. Quantum memory photon echo-like techniques in solids. *J. Opt. B: Quantum semiclass. Opt.* **5**, S497-S502 (2003).
3. Kraus. B, *et al.* Quantum memory for nonstationary light fields based on controlled reversible inhomogeneous broadening. *Phys. Rev. A* **73**, 020302 (2006).
4. Alexander, A. L. Longdell, J. J. Sellars, M. J. and Manson, N. B. Photon echoes produced by switching electric fields. *Phys. Rev. Lett.* **96**, 043602 (2006).
5. Hedges, M. P. Longdell, J. J. Li, Y. and Sellars, M. J. Efficient quantum memory for light. *Nature* **465**, 1052-1056 (2010).
6. Riedmatten, H. de *et al.* A solid-state light-matter interface at the single-photon level. *Nature* **456**, 773-777 (2008).
7. Saglamyurek, E. et al., Broadband waveguide quantum memory for entangled photons. *Nature* **469**, 512-515 (2011).
8. Afzelius, M. *et al.* Demonstration of atomic frequency comb memory for light with spin-wave storage. *Phys. Rev. Lett.* **104**, 040503 (2010).
9. Gündoğan, M. *et al.,* Coherent storage of temporally multimode light using a spin-wave atomic frequency comb memory. *New J. Phys.* **15**, 045012 (2013).
10. Damon, V. et al., Revival of silenced echo and quantum memory for light. *New. J. Phys.* **13**, 093031 (2011).
11. Dajczgewand, J. *et al*., Optical memory bandwidth and multiplexing capacity in the erbium telecommunication window. *New J. Phys.* **17**, 023031 (2015).
12. Arcangeli, A. Ferrier, A. and Goldner, Ph. Stark echo modulation for quantum memories. *Phys. Rev. A* **93**, 062303 (2016).
13. McAuslan, D. L. et al., Photon-echo quantum memories in inhomogeneously broadened two-level atoms. *Phys. Rev. A* **84**, 022309 (2011).
14. Ham, B. S. Atom phase controlled noise-free photon echoes. *arXiv*:1101.5480v2 (2011).
15. Ham, B. S. Coherent control of collective atom phase for ultralong, inversion-free photon echoes. *Phys. Rev. A 85, 031402(R) (2012); ibid, Phys. Rev. A* **94**, 049905(E) (2016).
16. Ham, B. S. A controlled ac Stark echo for quantum memories. Scientific Reports (To be published), *arXiv*:1612.02193 (2016).
17. Rahmatullah and Ham, B. S. "Analysis of Controlled Coherence Conversion in a Double Rephasing Scheme of Photon Echoes for Quantum Memories. *arXiv*: 1612.02167v2 (2016).

18. Ham, B. S. Control of photon storage time using phase locking. *Opt. Exp.* **18**, 1704 (2010).
19. Hahn, J. and Ham, B. S. Rephasing halted photon echoes using controlled optical deshelving. *New J. Phys.* **13**, 093011 (2011).
20. Ham, B. S. Gaussian beam profile effectiveness on double rephasing photon echoes. *arXiv*:1701.04291 (2017).
21. Steane, A. M. Efficient fault-tolerant quantum computing. *Nature* **399**, 124 (1999).
22. Lo, H.-K. Curty, M. and Qi, B. Measurement-Device-Independent Quantum Key Distribution. *Phys. Rev. Lett.* **108**, 130503 (2012).
23. Ham, B. S. "Collective atom phase controls in photon echoes for quantum memory applications I: population inversion removal," *arXiv*:1612.00115 (2016).
24. Ham, B. S., Shahriar, M. S., Kim, M. K., and Hemmer, P. R. Spin coherence excitation and rephasing with optically shelved atoms. *Phys. Rev. B* 58, R11825 (1998).
25. Sangouard, N. Simon, C. Afzelius, M. and Gisin, N. Analysis of a quantum memory for photons based on controlled reversible inhomogeneous broadening. *Phys. Rev. A* **75**, 032327 (2007).
26. Zhong M et al. Optical addressable nuclear spin in a solid with a six-hour coherence time *Nature* **517**, 177 (2015).

**Acknowledgment**
The present work was supported by the ICT R&D program of MSIP/IITP (1711028311: Reliable crypto-system standards and core technology development for secure quantum key distribution network).

**Supplemental Material** for "Understanding of collective atom phase control in modified photon echoes for a near perfect, storage time extended quantum memory," by Rahmatullah and B.S. Ham

We consider a Λ-type three-level optical medium whose upper-state is $|2\rangle$ and two lower states are $|1\rangle$ and $|3\rangle$ as shown in Fig. S1(a). The data pulse D and the rephasing pulse $R_1$ are coupled to transition of $|1\rangle - |2\rangle$, satisfying the two-pulse photon echo scheme. The two counter-propagating $C_1$ and $C_2$ control pulses are coupled to the transition of $|2\rangle - |3\rangle$. The optical Bloch and the Maxwell-Schrödinger equations are:

$$\frac{\partial}{\partial t}\sigma_{11}(z,t) = i\varepsilon_l(z,t)\big(\sigma_{12}(z,t) - \sigma_{21}(z,t)\big), \tag{S1}$$

$$\frac{\partial}{\partial t}\sigma_{22}(z,t) = i\varepsilon_l(z,t)\big(\sigma_{21}(z,t) - \sigma_{12}(z,t)\big) + i\varepsilon_j\big(\sigma_{23}(z,t) - \sigma_{32}(z,t)\big), \tag{S2}$$

$$\frac{\partial}{\partial t}\sigma_{33}(z,t) = i\varepsilon_j\big(\sigma_{32}(z,t) - \sigma_{23}(z,t)\big), \tag{S3}$$

$$\frac{\partial}{\partial t}\sigma_{12}(z,t) = i\Delta\sigma_{12}(z,t) + i\varepsilon_l(z,t)\big(\sigma_{11}(z,t) - \sigma_{22}(z,t)\big) + i\varepsilon_j(z,t)\sigma_{13}(z,t), \tag{S4}$$

$$\frac{\partial}{\partial t}\sigma_{32}(z,t) = i\Delta\sigma_{32}(z,t) + i\varepsilon_j(z,t)\big(\sigma_{33}(z,t) - \sigma_{22}(z,t)\big) + i\varepsilon_l(z,t)\sigma_{31}(z,t), \tag{S5}$$

$$\frac{\partial}{\partial t}\sigma_{13}(z,t) = i\varepsilon_j(z,t)\sigma_{12}(z,t) + i\varepsilon_l\sigma_{23}(z,t), \tag{S6}$$

$$\frac{\partial}{\partial z}\varepsilon_{l(j)}(z,t) = \frac{i\alpha}{2\pi}\int_{-\infty}^{\infty}\sigma_{12}(z,t)d\Delta, \tag{S7}$$

where, $l$= D or $R_1$ and $j$= $C_1$ or $C_2$. In the following, we consider the detailed calculations for a single rephasing scheme in a three-level atom. The corresponding pulse sequence is shown in Fig. S1(b).

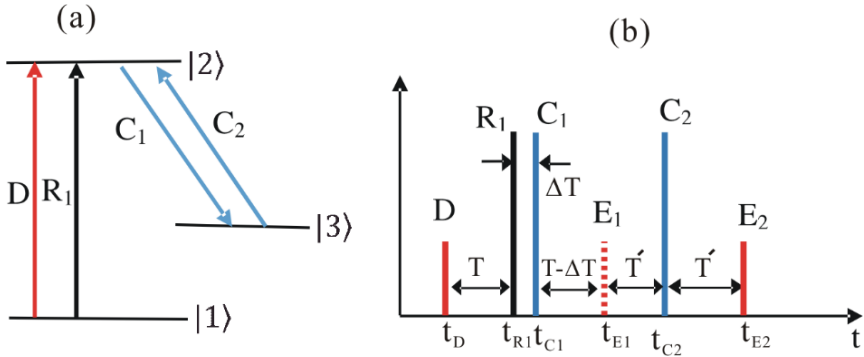

**Figure S1.** A schematic of single rephasing photon-echo based quantum memory protocol in a three-level system. (a) Energy level diagram and (b) Pulse sequence for (a). $t_j$ is the arrival time of pulse j.

**D-pulse**

A weak D-pulse propagates through the medium along z-direction. We assume that D-pulse is much smaller than π-pulse, so that nearly all atoms remain in the ground state, i.e., $\sigma_{11}(z,t) = 1$. The resultant Maxwell-Bloch equations are obtained by putting $\varepsilon_l = \varepsilon_D$ and $\varepsilon_j = 0$ in equations (S4) and (S7)

$$\frac{\partial}{\partial t}\sigma_{12}(z,t,\Delta) = i\Delta\sigma_{12}(z,t) + i\varepsilon_D(z,t), \tag{S8}$$

$$\frac{\partial}{\partial z}\varepsilon_D(z,t) = \frac{i\alpha}{2\pi}\int_{-\infty}^{\infty}\sigma_{12}(z,t)d\Delta. \tag{S9}$$

The equation (S8) is first order linear differential equation with the following solution:

$$\sigma_{12}(z,t,\Delta) = i\int_{-\infty}^{t}\varepsilon_D(z,\acute{t})e^{i\Delta(t-\acute{t})}d\acute{t}. \tag{S10}$$

Initially the atomic coherence is zero, $\sigma_{12}(z,-\infty) = 0$. In equation (S10), the positive sign of coherence stands for absorption. Taking the Fourier transform of equation (S8) and substituting it into the Fourier version of (S9), we obtain:

$$\frac{\partial}{\partial z}\varepsilon_D(z,\omega) = -\frac{\alpha}{2\pi}\int_{-\infty}^{\infty}\varepsilon_D(z,\omega)\left[\frac{1}{i(\omega-\Delta)} + \pi\delta(\omega-\Delta)\right]d\Delta = -\frac{\alpha}{2}\varepsilon_D(z,\omega). \tag{S11}$$

The solution of equation (S11) in time domain is

$$\varepsilon_D(z,t) = e^{-\frac{\alpha z}{2}}\varepsilon_D(0,t). \tag{S12}$$

The D-pulse exponentially decays and absorbed by the medium.

### R$_1$-pulse

In time T after the D-pulse, we apply a $\pi$ R$_1$-pulse propagating in z-direction to retrieve the absorbed D information via rephasing process. After R$_1$-pulse the populations between states |1⟩ and |2⟩ are swapped to be $\sigma_{22}(z,t) = 1$, and the corresponding equations of motion are:

$$\frac{\partial}{\partial t}\sigma_{12}(z,t,\Delta) = i\Delta\sigma_{12}(z,t) - i\varepsilon_{R_1}(z,t), \tag{S13}$$

$$\frac{\partial}{\partial z}\varepsilon_{R_1}(z,t) = \frac{i\alpha}{2\pi}\int_{-\infty}^{\infty}\sigma_{12}(z,t)d\Delta. \tag{S14}$$

The equations (S13) and (S14) are obtained by substituting $\varepsilon_l = \varepsilon_{R_1}$ and $\varepsilon_j = 0$ in equations (S4) and (S7). The solution of equation (S13) yields:

$$\sigma_{12}(z,t,\Delta) = e^{i\Delta(t-t_{R_1})}\sigma_{12}(z,t_{R_1}) - i\int_{t_{R_1}}^{t}\varepsilon_{R_1}(z,\acute{t})e^{i\Delta(t-\acute{t})}d\acute{t}. \tag{S15}$$

Due to the R$_1$ reaphsing the coherence $\sigma_{12}$ at $t = t_{R_1}$ is equal to the Hermitian conjugate of equation (S10).

$$\sigma_{12}(z,t,\Delta) = -ie^{-i\Delta(2t_{R_1}-t)}\int_{-\infty}^{\infty}\varepsilon_D^{\dagger}(z,\acute{t})e^{i\Delta\acute{t}}d\acute{t} - i\int_{t_{R_1}}^{t}\varepsilon_{R_1}(z,\acute{t})e^{i\Delta(t-\acute{t})}d\acute{t}. \tag{S16}$$

The negative sign shows the emissive coherence. After doing some mathematical treatments, the following equation is obtained in a frequency domain:

$$\sigma_{12}(z,\omega,\Delta) = -ie^{-2i\Delta t_{R_1}}2\pi\delta(\omega-\Delta)\int_{-\infty}^{\infty}\varepsilon_D^{\dagger}(z,\acute{t})e^{i\Delta\acute{t}}d\acute{t} - i\varepsilon_{R_1}(z,\omega)\left[\frac{1}{i(\omega-\Delta)} + \pi\delta(\omega-\Delta)\right]. \tag{S17}$$

Applying Fourier transform to equation (S14) and substituting equation (S17), we get:

$$\varepsilon_{R_1}(z,t) = \varepsilon_{R_1}(0,t)e^{\frac{\alpha z}{2}} + 2\sinh\left(\frac{\alpha z}{2}\right)\varepsilon_D^\dagger(0, 2t_{R_1}-t). \tag{S18}$$

The echo is emitted at $t = 2t_{R_1} - t_D$. The efficiency of the echo is $4\sinh^2\left(\frac{\alpha z}{2}\right)$, where the echo intensity is greater than unity for large optical depth. This echo amplification is simply due to the inverted medium.

### $C_1$-pulse

The function of $C_1$–pulse is to transfer the optical coherence $\sigma_{12}$ into spin coherence $\sigma_{13}$ via complete population transfer from the optical state $|2\rangle$ to the spin state $|3\rangle$. The pulse $C_1$ is delayed by time $\Delta T$ from $R_1$ and the coherence at $t = t_{C1}$ is given by:

$$\sigma_{12}(z, t_{C1}) = -i\,e^{-i\Delta(2t_{R_1}-t_{C1})} \int_{-\infty}^{\infty} \varepsilon_D^\dagger(z,\acute{t})e^{i\Delta \acute{t}}\,d\acute{t}. \tag{S19}$$

We only consider the first part of equation (S16) related to the D-pulse. The optical Bloch equations after $C_1$ by putting $\varepsilon_l = 0$ and $\varepsilon_j = \varepsilon_{C_1}$ in equations (S4) and (S6) are as follow:

$$\frac{\partial}{\partial t}\sigma_{12}(z,t) = i\Delta\sigma_{12}(z,t) + i\varepsilon_{C_1}(z,t)\sigma_{13}(z,t), \tag{S20}$$

$$\frac{\partial}{\partial t}\sigma_{13}(z,t) = i\varepsilon_{C_1}(z,t)\sigma_{12}(z,t). \tag{S21}$$

After solving equations (S20) and (S21), we get $\sigma_{12}(z,t) = 0$ and $\sigma_{13}(z,t) \neq 0$. Thus, the initial optical coherence is locked unless there is spin dephasing for $\sigma_{13}$.

### $C_2$-pulse

The pulse $C_2$ transfers the spin coherence back to the optical coherence. By substituting $\varepsilon_l = 0$ and $\varepsilon_j = \varepsilon_{C_2}$ and $\sigma_{13}(z,t) = 0$ in equation (S4), we obtain:

$$\frac{\partial}{\partial t}\sigma_{12}(z,t) = i\Delta\sigma_{12}(z,t). \tag{S22}$$

The solution of the above equation is

$$\sigma_{12}(z,t,\Delta) = \sigma_{12}(z,t_{C_2})e^{i\Delta(t-t_{C_2})}. \tag{S23}$$

The coherence at $t_{C_2}$ is $\sigma_{12}(z,t_{C_2}) = -\sigma_{12}(z,t_{C_1})$, where the negative sign is due to the phase shift induced by both C1 and C2. The equation (S23) can be written as:

$$\sigma_{12}(z,t) = i\,e^{-i\Delta(2t_{R_1}-t_{C_1}+t_{C_2}-t)} \int_{-\infty}^{\infty} \varepsilon_D^\dagger(z,\acute{t})e^{i\Delta \acute{t}}\,d\acute{t}. \tag{S24}$$

The final atomic coherence is positive, in which the echo is absorptive and cannot be radiated from the medium. If the area of the $C_2$ is increased to $3\pi$, then $\sigma_{12}(z,t) = -\sigma_{12}(z,t)$, where the echo is emissive. Therefore, $\pi - 3\pi$ pulse sequence of $C_1$ and $C_2$ is a valid condition.